\documentclass{amsart} 
\usepackage{mathrsfs,amssymb,amsmath,amsxtra,amsfonts}
\usepackage{graphicx,color}
\usepackage{verbatim}
\usepackage[colorlinks]{hyperref}

\begin{document}

\title[Black holes or frozen stars?]{Black holes or frozen stars? A viable theory of gravity without black holes} \sloppypar\sloppy

\author{I. Schmelzer}
\thanks{Berlin, Germany}
\email{ilja.schmelzer@gmail.com}%
\urladdr{ilja-schmelzer.de}

\begin{abstract}
Do observations of black hole candidates rule out alternative theories of gravity without horizon formation? This depends on the existence, viability and reasonableness of alternative theories of gravity without black holes. 

Here a theory of gravity without black hole horizon formation is presented. The gravitational collapse stops shortly before horizon formation and leaves a stable frozen star. In the limit $\Xi, \Upsilon\to 0$ the Einstein equations of GR are recovered, and the frozen stars become observationally indistinguishable from GR black holes. The theory therefore provides a counterexample to recent claims that observational evidence from black hole candidates ``all but requires the existence of a horizon''.

The theory presented here shares its equations with RTG. Nonetheless, as is shown, there remain important conceptual and physical differences. In particular, some serious problems of RTG are not present in the theory proposed here. So it can be argued that the theory is a physically viable and conceptually sound alternative to GR.
\end{abstract}

\maketitle

\newcommand{\x}{\mbox{$\mathfrak{x}$}}
\renewcommand{\t}{\mbox{$\mathfrak{t}$}}
\renewcommand{\a}{\alpha}
\renewcommand{\b}{\beta}
\newcommand{\pd}{\mbox{$\partial$}} 
\providecommand{\abs}[1]{\lvert#1\rvert}
\newcommand{\R}{\mbox{$\mathbb{R}$}}

\newtheorem{theorem}{Theorem}
\newtheorem{axiom}{Axiom}
\newtheorem{postulate}{Postulate}
\newtheorem{definition}{Definition}
\newtheorem{thesis}{Thesis}
\newtheorem{corollary}{Corollary}

\section{Motivation}\label{motivation}

Broderick, Loeb and Narayan \cite{BN3} have claimed that ``recent millimeter and infrared observations of Sagittarius A* (Sgr A*), the supermassive black hole at the center of the Milky Way, all but requires the existence of a horizon''. Moreover, they specify that they ``are able to frame this argument entirely in terms of observable quantities'', and that ``our results apply to all geometric theories of gravity that admit stationary solutions''.

Is this claim correct? The aim of this text is to present a counterexample -- a metric theory of gravity which admits, if one of the free paramters of the theory $\Upsilon$ is greater zero, stationary solutions.  But in the limit $\Upsilon \to 0$ these stationary solutions become indistinguishable from a GR black hole solution if observation is restricted to outside observers. 

The theory presented here as a counterexample has been defined in \cite{Schmelzer,Schmelzer1} and will be named ``general Lorentz ether theory'' (GLET). The Lagrangian, and, therefore, the equations of motion are, for special conditions on the free parameters of the theory $\Xi, \Upsilon>0$ and for Einstein's cosmological constant $\Lambda<0$, equivalent to those of the relativistic theory of gravity (RTG) proposed by Logunov and coworkers \cite{Logunov2}. The results presented here for black holes in GLET for this parameter constellation are, therefore, applicable to RTG too, and one may think that GLET is only an irrelevant metaphysical interpretation of RTG. But the difference in the seemingly only metaphysical interpretation has an important influence on the physics, in form of different ``causality conditions'' in above theories. Now, the RTG causality condition is a quite problematic thing, because it is incompatible with the equations of motion: Even if it is fulfilled for some initial conditions, it may be violated later, thus, the whole theory seems conceptually inconsistent.

The metaphysics of GLET does not lead to such conceptual problems. While, for the different causality condition of GLET, a similar situation is possible too, it obtains a metaphysically meaningful interpretation: The equations of any continuous condensed matter theory becomes physically invalid at places where the density becomes zero and the material tears into parts. With this much better justification of the causality condition in GLET, another problem of RTG -- the ghost instability -- can be solved too. Last but not least, RTG requires a seemingly wrong sign $\Lambda<0$ for Einstein's cosmological constant, thus, needs some dark energy or quintessence to describe an accelerating expansion of the universe, while there is no such restriction in GLET. 

Thus, the most powerful known arguments against RTG are not applicable against GLET. So it becomes a viable alternative to GR, devoid of the problems of RTG as well as the problems of GR with causality protection and singularities. 

But is it only an unexpected exception that GLET leads to a modification of GR only in a region extremely close to the GR horizon? We argue that it is not, that it is, instead, very likely that many reasonable alternative theories of gravity will show similar behaviour. The argument we propose is metaphysical in its nature: What we experience as the flow of time is not adequately described by actual GR spacetime metaphysics. If one corrects this fault, one easily ends up with theories where GR is corrected by small non-covariant terms, which allow to define a preferred time. Such theories are, then, likely to differ from GR only very close to the horizon, because here a very natural candidate for a preferred time coordinate, harmonic time, becomes singular. 

So, Broderick et al deserve praise for looking carefully at one of the domains where it is quite likely that GR may be empirically falsified, where a large class of reasonable alternatives to GR will make different empirical predictions. 

\section{Definition of the theory}\label{sec:definition}

The theory of gravity defined in \cite{Schmelzer,Schmelzer1}, which is considered here, is defined in the following way: Similar to GR and other metric theories of gravity, we have the gravitational field $g_{\mu\nu}(x)$ together with some unspecified number of matter fields $\varphi_m(x)$.  Different from GR, these fields are not defined on an arbitrary four-dimensional manifold, but on a classical Newtonian spacetime $\R^3\times\R$, with a set of preferred background coordinates $\x^\a$, in particular a preferred time $\x^0=\t$. Latin indices $i,j,\ldots$ refer to the preferred spatial coordinates $\x^i$, $1\le i\le 3$, lower greek indices  $\a,\b,\ldots$ refer to preferred four-coordinates $\x^\alpha$, $0\le\alpha\le 3$, while upper greek indices $\mu,\nu,\ldots$ refer to general spacetime coordinates $x^\mu$, $0\le\mu\le 3$. 

In the preferred coordinates $\x^\a$, the action of the theory is
\begin{equation}\label{S}
S = -\frac{1}{16\pi G}\int
       	(R + 2\Lambda + \Upsilon g^{00}-\Xi g^{ii})\sqrt{-g}d^4\x
         +   \int L_{matter}\sqrt{-g}d^4\x\\
\end{equation}
where $L_{matter}$ a covariant Lagrangian for the matter fields. Introducing the diagonal matrix $(\gamma_{\alpha\beta}) = \text{diag}(\Upsilon,-\Xi,-\Xi,-\Xi)$, one can rewrite this as
\begin{equation}\label{Sgamma}
S = -\frac{1}{16\pi G}\int
       	(R + 2\Lambda + \gamma_{\a\b}g^{\a\b})\sqrt{-g}d^4\x
         +   \int L_{matter}\sqrt{-g}d^4\x\\
\end{equation}
or, in a covariant form, where the preferred coordinates appear as functions $\x^\a(x)$ of general coordinates $x$, as
\begin{equation}\label{Scovgamma}
S = -\frac{1}{16\pi G}\int
        (R + 2\Lambda + \gamma_{\a\b}g^{\mu\nu}\x^\a_{,\mu}\x^\b_{,\nu})\sqrt{-g}d^4x
         +   \int L_{matter}\sqrt{-g}d^4x\\
\end{equation}
We also need -- for reasons which become clear below -- the additional ``causality condition'' that the preferred time $\t=\x^0$ has to be time-like everywhere. With this additional condition, the definition of the theory is already complete.

The Euler-Lagrange equations for the $g_{\a\b}$ give
\begin{equation}\label{GLETeq}
G^\a_\b  = 8\pi G (T_m)^\a_\b
   + (\Lambda +\frac12\gamma_{\gamma\delta}g^{\gamma\delta}) \delta^\a_\b
   - g^{\a\gamma}\gamma_{\gamma\b}.
\end{equation}
while variation over the $\x^\a$ in \eqref{Scovgamma} gives the harmonic equations
\begin{equation}\label{harm}
\partial_\a  (g^{\a\b}\sqrt{-g}) =  \square_g \x^\b = 0,
\end{equation}
which are also a consequence of the equations \eqref{GLETeq}.
The harmonic equation has (modulo constants) the form of a conservation law
\begin{equation}
\partial_\a  \mathcal{T}^\a_\b  = 0
\end{equation}
with an (densitized) energy-momentum tensor defined by
\begin{equation}
\mathcal{T}^\a_\b = \frac{1}{8\pi G}\gamma_{\b\gamma}g^{\a\gamma}\sqrt{-g}
\end{equation} 
The equations of motion \eqref{GLETeq} allow to split this energy-momentum tensor into a ``matter part''
\begin{equation}
(\mathcal{T}_m)^\a_\b  = (T_m)^\a_\b\sqrt{-g} 
\end{equation}  and a ``gravitational part''
\begin{equation}
(\mathcal{T}_g)^\a_\b = \frac{1}{8\pi G}\left(\delta^\a_\b
			(\Lambda + \frac12\gamma_{\gamma\delta}g^{\gamma\delta})
              - G^\a_\b\right)\sqrt{-g}.
\end{equation}
so that 
\begin{equation}
\mathcal{T}^\a_\b  = (\mathcal{T}_m)^\a_\b + (\mathcal{T}_g)^\a_\b
\end{equation} 

\section{The condensed matter interpretation}\label{sec:interpretation}

The theory allows a condensed matter interpretation. In a variant of the ADM decomposition \cite{ADM} we can use the preferred coordinates $\x^i$, $\t$, to split the gravitational field $g_{\mu\nu}$ into a scalar $\rho$, a three-vector $v^i$ and a three-metric $\sigma^{ij}$:
\begin{subequations}\label{gdef}
\begin{align}
g^{00}\sqrt{-g} &= \rho, \\
g^{0i}\sqrt{-g} &= \rho v^i, \\
g^{ij}\sqrt{-g} &= \rho v^i v^j - \sigma^{ij}.
\end{align}
\end{subequations}
Their condensed matter interpretation becomes obvious if we look what happens with the harmonic equations \eqref{harm}. They become
\begin{subequations}\label{classical}
\begin{align}
\label{continuity}
\partial_t \rho + \partial_i (\rho v^i) &= 0, \\
\label{Euler}
\partial_t (\rho v^j) + \partial_i(\rho v^i v^j - \sigma^{ij}) &= 0.
\end{align}
\end{subequations}
These are the continuity and Euler equations well-known from classical condensed matter theory if we identify $\rho$ with the density of some fundamental medium, $v^i$ with its velocity, and $\sigma^{ij}$ with its stress tensor.

What about the matter fields? If there would be other media interacting with our medium, this would require some momentum exchange, thus, some additional interaction terms in the Euler equation \eqref{Euler}. Now the equations \eqref{classical} hold also in the case of interaction with matter fields. But there is a simple solution of this problem: The matter fields have to be interpreted as fields which describe some other \emph{material properties of the same medium} instead of properties of other media.

Of course, to give a condensed matter interpretation of a metric theory of gravity raises the problem of explanation of relativistic symmetry. Indeed, situations which are equivalent from point of view of the space-time interpretation are described, depending on the particular choice of the time coordinate $\t$, in a physically different way, in particular as configurations with different $\rho$, $v^i$, and $\sigma^{ij}$. But such an explanation of relativistic symmetry is possible -- the general Lagrangian can be derived from independent first principles. This derivation is given in appendix \ref{derivation}.

With this interpretation, the theory proposed here appears as a natural generalization of the classical Lorentz ether to gravity. Let's note that many classical arguments against the Lorentz ether are no longer applicable to this generalization: It has a Lagrange formalism, thus, has the resulting ``action equals reaction'' symmetry, it is, in particular, no longer incompressible, and  it provides an explanation of relativistic symmetry. Then it no longer describes only electromagnetism, but all fields including gravity. 

Thus, let's name it ``general Lorentz ether theory'' (GLET).

Of course, the theory describes, essentially, only the gravitational field. All other fields are only characterized as other, unspecified ``material properties'' of the ether. In this sense, the theory is quite incomplete. It is only a general theory, for a whole class of more specific models, where the material properties have to be specified in detail. 

Such a proposal for a particular model of the material properties, which gives the fermions and gauge fields of the standard model of particle physics, and which is at least conceptually compatible with this theory, has been proposed in \cite{clm}. 

\section{Comparison with the Relativistic Theory of Gravity (RTG)}\label{sec:RTG}

The Lagrangian \eqref{Sgamma} of the theory itself is not new. For the particular choice of the signs of the free parameters as $\Xi,\Upsilon>0$ and $\Lambda<0$, it is the Lagrangian of the ``relativistic theory of gravity'' (RTG). This theory has been first proposed by \mbox{Freund}, Maheshwari and Schonberg 1969 \cite{FMS} and appears in a two-parameter family of Lagrangians constructed by Ogievetsky and Polubarinov 1965 \cite{OP} as the case $q=0$, $p=-1$.  Later, the Lagrangian has been rediscovered by Logunov and coworkers \cite{Logunov,Logunov1,Logunov2,Logunov3}, who have named the theory ``relativistic theory of gravity'' (RTG).  Because this name is widely used (see \cite{GershteinFlow}--\cite{NieuwenhuizenBH}) and neutral we will use it too (instead of, say, FMS theory). For an introduction into RTG see \cite{Logunov2}.

RTG is a bimetric theory. It contains, together with an effective physical metric $g_{\mu\nu}(x)$, also a fixed Minkowski background metric $\eta^{\mu\nu}$, which is, at the same time, the unique vacuum solution of the theory. The gravitational field is a spin $2$ field $h_{\mu\nu}$ which moves on this Minkowski background. It uniquely defines and is uniquely defined by the effective metric $g_{\mu\nu}(x)$ by the formula
\begin{equation}
h^{\mu\nu}=g^{\mu\nu}\sqrt{-g}-\eta^{\mu\nu}\sqrt{-\eta}.
\end{equation}
Conceptually all fields, the gravitational field $h_{\mu\nu}$ as well as the matter fields, move on the Minkowski background $\eta_{\mu\nu}(x)$. Gravitational interaction leads to an effective replacement of $\eta_{\mu\nu}(x)$  by the effective metric $g_{\mu\nu}(x)$. But this gravitational interaction cannot violate the Einstein causality of the background metric $\eta_{\mu\nu}(x)$. As a consequence, the light cone of $g_{\mu\nu}(x)$ should be inside the light cone of the background $\eta_{\mu\nu}(x)$. This is an additional ``causality condition'', which does not follow from the equations of motion of RTG. 

Gravity in RTG is massive. The connection between the graviton mass $m>0$, the background metric $\eta^{\mu\nu}$, and the parameters $\Upsilon, \Xi, \Lambda$ of the theory presented here is defined by 
\begin{equation}
        \Lambda=-\frac{m^2}{ 2},\qquad
        \Xi=-\eta^{11}\frac{m^2}{ 2},\qquad
        \Upsilon=\eta^{00}\frac{m^2}{ 2}.
\end{equation} 
This implies $\Upsilon>0, \Xi>0, \Lambda<0$. For this choice of the parameters of GLET, we obtain an exact correspondence between the Lagrangian, the equations of motion and the solutions of above theories. This exact correspondence can and will be used to take over mathematical results about RTG solutions to our theory. 

In particular, we obtain for these choices of the parameters a constant (vacuum) solution 
\begin{equation}
g^{\mu\nu}_{vac} = \eta^{\mu\nu}= -\Lambda^{-1}\gamma^{\mu\nu}.
\end{equation}
It also follows that the equation \eqref{GLETeq} reduces in the Newtonian limit around $g^{\mu\nu}_{vac}$ to
\begin{equation}
 (\Delta-m^2) V = \kappa \rho_{matter},
\end{equation} 
so that the action \eqref{Sgamma} defines a bimetric theory on a Minkowski background with massive graviton. This massive gravitational field ``has spins $2$ and $0$ and represents a physical field in the Faraday-Maxwell spirit'' \cite{Logunov2}. 

The relation between GLET and RTG is therefore similar to the relation between the Lorentz ether interpretation and the Minkowski spacetime interpretation of special relativity. As far as we consider only the Lagrangian and the equation, and restrict ourself to the choice of signs $\Upsilon>0$, $\Xi>0$, $\Lambda<0$, there is indeed a complete equivalence between the equations of the two theories.

Now, RTG is not a popular theory of gravity. There are some serious arguments against this theory. Given that the Lorentz ether interpretation of special relativity is also quite unpopular in comparison with the Minkowski spacetime interpretation, one may ask why one should consider an unpopular metaphysical interpretation of an already unpopular theory of gravity. 

The reason is that the equivalence between the theories is not complete, and that these remaining differences become important if we evaluate the popular arguments against RTG -- they become invalid as arguments against GLET.

\subsection{Different causality conditions}   

So let's see where the similarity is not complete. The two different interpretations of the effective metric $g_{\mu\nu}$ -- in our theory in terms of $\rho,v^i,\sigma^{ij}$, in RTG in terms of $\eta_{\mu\nu},h_{\mu\nu}$ -- have a different domain of applicability, which leads to different ``causality conditions'' for above interpretations. In GLET, the condition is that the density $\rho(\x)$ has to be non-negative. In RTG, the light cone of the effective metric $g_{\mu\nu}$ has to be inside the light cone of the fundamental metric $\eta_{\mu\nu}$ of RTG. 

Above causality conditions are important for the meaningfulness of the metaphysical interpretations. And, in particular, above conditions enforce something which is missed in general relativity, namely chronology protection: A metric which fulfills the causality conditions is a global hyperbolic metric, thus, it does not have any closed time-like curves (CTCs). The conceptual possibility of such CTCs is probably the most important metaphysical problem of the spacetime interpretation of GR. Thus, the causality conditions are conceptually important restrictions, which serve a purpose. 

Unfortunately, above causality conditions appear to be in conflict with the evolution equations of the theory. The additional terms in the equations do not define a mechanism for chronology protection. In particular, it is easy to find solutions of RTG where the causality condition is fulfilled for the initial conditions but becomes violated later. Such solutions exist even arbitrary close to the vacuum solution.  And every RTG solution of this type also allows to define GLET solutions which violate the GLET causality condition. 

But there exists an important metaphysical difference between violations of the RTG causality condition, interpreted from point of view of RTG metaphysics, and violations of the GLET causality condition interpreted from point of view of the condensed matter interpretation.  

Namely, the RTG metaphysics leaves this situation completely unexplained and unexplainable. Something has to be wrong with these solutions, they cannot be solutions of RTG. But what is wrong with them?  What is, in particular, wrong with the initial conditions, which do not violate the causality condition? There should be something wrong with them -- once they do not lead to valid solutions of the theory. Or there should be something wrong with the equations of RTG -- they should be modified in the critical region where the violation of the RTG causality condition begins.  

Surprisingly, the metaphysical situation is much better in GLET. Here, the causality condition is violated if the density of the Lorentz ether becomes zero. Of course, this means that the condensed matter theory can no longer be applied.  In particular, regions of the solution where the density is below zero are clearly physically meaningless. 

But this is only a problem of the particular equations for the Lorentz ether. A condensed matter interpretation of some set of equations presupposes that the equations become invalid if the density becomes too small. So, the mathematics of the theory has to be changed in the regions with too small or zero density of the ether. This is natural from the very point of view of the interpretation, which assumes from the start that the condensed matter equations are only approximations of some different, more fundamental theory. 

Nothing comparable can be said about RTG. Nothing in the theory and its interpretation suggests that the equations are only approximations of some other, different theory, and no idea is given about the nature of such a more fundamental theory. 

\subsection{Restrictions for global cosmology}

Together with the Einstein-Hilbert Lagrangian of general relativity, Einstein's cosmological term is also part of above theories. But RTG imposes an additional restriction to this term: Einstein's cosmological constant has to be negative> $\Lambda < 0$. 

As a consequence, RTG does not predict an acceleration of the expansion of the universe, as it seems to follow from observation. Thus, RTG needs some strange form of matter, called quintessence, to explain accelerated expansion.  

This restriction is not present in GLET. The interpretation, as well as the derivation of the Lagrangian, do not contain anything which would fix the sign of $\Lambda$. 

Therefore the observation of $\Lambda$ as being positive defines a strong empirical problem for RTG, but not for GLET. 

Another, more subtle difference, which may nonetheless be important for cosmology, is the possible value of $\Xi$. Large enough values of $\Xi$ do not allow for an accelerating expansion, but can make expansion close to constant: $a(\tau)\approx a_0\tau$ (also called ``coasting''). Without it, with usual (even dark) matter only, we would have a deceleration of the expansion. Now, alternative explanations of the empirical evidence for accelerated expansion seem at least possible (see, for example, the explanation proposed by Wiltshire \cite{Wiltshire}). But, even if they fail, it may be that a ``freely coasting'' universe may remain viable (see \cite{Lohiya}), while a decelerating universe without any additional terms fails.  

In this case, the difference between the causality conditions of RTG and GLET may become important too. In RTG, the value for $\Xi$ has to be chosen in such a way that the background metric always contains the physical metric, which is, in RTG, oscillating. As a consequence, $\Xi$ has to be so small that it does not lead to any observable effects. 

The situation is different in GLET, where the causality condition does not impose any conditions on $\Xi$. So it can be chosen large enough to have an influence on global cosmology, to enforce a freely coasting universe instead of one with decelerating expansion. 

\subsection{The problem with the ghost instability}

But there is also another popular argument against RTG, which is part of a general argument against massive theories of gravity. In the linear approximation, RTG, and, therefore, also GLET (for $\Xi,\Upsilon>0,\Lambda<0$), is such a massive theory of gravity.  

At a first, naive look, there seems to be no problem to obtain GR in the natural limit $\Xi,\Upsilon\to 0$. Unfortunately it requires a more detailed consideration, because in the case of massive gravity with the Fierz-Pauli mass term it seems equally obvious but is wrong --  it leads to the van Dam-Veltman-Zakharov discontinuity \cite{vanDam1},\cite{vanDam2},\cite{ZakharovDisc},\cite{Boulware}. In \cite{Grishchuk02}, the discontinuity is understood as a consequence of an infinite mass for the spin-0 component, which does not disappear if the mass for the spin-2 component becomes zero. The RTG mass term is not of Fierz-Pauli type, so that with $m\to 0$ the masses of above spins go to zero and no discontinuity appears.

But there was a reason for considering the Fierz-Pauli mass term. Before the discontinuity has been found, it was believed to be the only viable mass term: All other mass terms lead, in the linear approximation, to tachyons or ghosts. And, indeed, the linear approximation of RTG contains a ghost because of the spin-0 component. Ghosts are supposed to lead to negative energies and instabilities, therefore there is a strong prejudice against them. 

Fortunately, it has been shown by Loskutov \cite{Loskutov} that the flow of gravitational energy from an isolated sourse is positive-definite (see also \cite{GershteinFlow}). Grishchuk, who has heavily criticized RTG in \cite{Grishchuk88}, \cite{Grishchuk90} and mentioned these problems as part of his critique in \cite{Grishchuk90}, has later demonstrated (together with Babak) ``the incorrectness of the widely held belief that the non-Fierz-Pauli theories possess negative energies and instabilities'' \cite{Grishchuk02}.

But is the situation, therefore, completely satisfactory? There may be some doubt, because \cite{Loskutov} considers only a particular situation, and, moreover, uses approximations. On the one hand, the problem has been formulated based on a simple, linearized approximation. So if the solution is presented in a similar form, using an approximation too, there seems not much to object. Nonetheless, some feeling of uncertainty remains. 

Fortunately, the possibility of an instability may be considered also from a global point of view, and independent of the particular situation. So, let's consider a particular solution, and let's denote the preferred coordinates of this solution as $x$, so that the solution is $g_{\a\b}(x)$. Now, assume there is an instability, so that a small difference in the initial values of the preferred coordinate leads to some other solution $\x^\a$. Now, if we look at the Lagrangian:
\begin{equation} 
S = -\frac{1}{16\pi G}\int
       	(R + 2\Lambda + \Upsilon g^{\mu\nu}\t_{,\mu}\t_{,\nu}-\Xi g^{\mu\nu}\x^i_{,\mu}\x^i_{,\nu})\sqrt{-g}d^4\x
         +   \int L_{matter}\sqrt{-g}d^4\x\\
\end{equation}
it looks locally like that of GR together with scalar fields $\t(x)$, $\x^i(x)$. And these scalar fields would have the usual properties of scalar fields for $\Xi>0$, $\Upsilon<0$. But the choice which corresponds to RTG is $\Upsilon>0$. So, the sign of the $\t(x)$ field seems to be the wrong one, with negative energy corresponding to large values of the $\t(x)$ field.

So, it seems, if we consider quantum fluctuations, we have to be afraid of instabilities where the $\t(x)$ field becomes infinite. 

But let's look how dangerous it is. Let's consider only fluctuations of the ``dangerous'' field -- the time coordinate, assuming $\x^i(x)=x^i$. Let's also ignore the back-reaction of the coordinates on the metric. Is it possible that $\t(x)$, initially close to $x^0$, increases to infinity for some finite $t=x^0$? Not. Near the infinity, the value of $g^{\a\b}\t_{,\a}\t_{,\b}=g^{00}\frac{\pd\t}{\pd x^0}^2$ becomes large in comparison with $g^{00}$, and it would have to become infinite if $\t$ becomes infinite. But that means that the conserved energy density $\rho=g^{00}\sqrt{g}$ would have to become infinite at $t$.

Other types of instabilities also seem to be prevented by the theory itself. Heavy oscillations, for example, do not define valid systems of coordinates. In particular, $\t(t)$ cannot be a time-like coordinate similar to $t$ initially, always remain time-like together with $t$, and later decrease in $t$. This also prevents $\t\to -\infty$ in some future.

What is the key of this consideration? Usual energy conservation, together with the positivity of energy for usual matter, as well as for the spatial coordinates considered as scalar fields, prevent usual instabilities. Here, the only exception is the time $\t$, which formally looks like a field with negative energy. But this field appears restricted in itself, first by its geometric character -- it has to be a global time-like coordinate -- and then by the remarkable other form of conservation law: The conservation of the ether itself, as described by the continuity equation \eqref{continuity}, together with the positivity $\rho>0$ of the ether density.

So there is no reason to be afraid of ghost instabilities. Instead, the situation is much more satisfactory than in general relativity, where local energy and momentum conservation is undefined. 

\subsection{Summary}

So we have found that, in agreement with results of Loskutov \cite{Loskutov} and Grishchuk \cite{Grishchuk02}, that we should not be afraid of negative energies and instabilities in RTG as well as GLET. 

We have also found some serious weaknesses of RTG -- the conceptual inconsistency between the RTG equations and the RTG causality condition, and the restrictions of the RTG parameters which may appear problematic for global cosmology. Above problems do not appear in GLET.
 
The initial presentations of RTG have been criticized also for other reasons, which have no relevance for the questions considered here. In particular, the original version of RTG has been motivated with incorrect claims about general relativity, like an inability of GR to give definite predictions for gravitational phenomena \cite{Logunov}. This has been correctly criticized, for example, by Seldovitch and Grishchuk \cite{Grishchuk88}. This has, clearly, not increased the popularity of RTG, but is irrelevant for the problems considered here. 

So we can conclude that the comparison with RTG, in particular the evaluation if the known arguments against RTG can be applied against GLET too, has not given any reasons to doubt the viability of GLET.

\section{Frozen stars instead of black holes}\label{sec:blackHoles}

On the other hand, given that the equations of above theories are identical, we can reuse now the results of RTG about black holes.

The gravitational collapse and its final state -- a black hole in GR, but something completely different in GLET -- is a domain where the additional terms in the GLET Lagrangian becomes important. Moreover, it remains important even for arbitrary small values of $\Xi$, $\Upsilon$. All we need is that $\Upsilon>0$.  

This is a surprising fact, which deserves a detailed consideration, and it is surprising for several reasons. 

First of all, the additional terms in the GLET Lagrangian do not depend on derivatives of the metric. This makes them similar to Einstein's cosmological term. But, in this case, why not apply the same reasoning, which is applicable to Einstein's cosmological term, to these terms too? This would suggest that the additional terms become important only on cosmological distances: Terms which depend on derivatives of the metric become more important locally, where they may be large for small local variations, but less important on large distances.  

And, indeed, the additional terms are important for global cosmology. For a homogeneous metric, which is  
\begin{equation}\label{FRWharmonic}
ds^2 = a^6(t) dt^2 - \beta^4 a^2(t)(dx^2+dy^2+dz^2).
\end{equation}
in harmonic time, the use of proper time $d\tau = a^3 dt$ gives the usual FRW-ansatz for the flat universe with some scaling factor $\beta>0$:
\begin{equation} ds^2 = d\tau^2 - \beta^4a^2(\tau)(dx^2+dy^2+dz^2). \end{equation}
This leads to ($p=k\varepsilon$, $8\pi G = c = 1$):
\begin{subequations}\label{eq:Friedman}
\begin{align}
\label{eq:Friedman1}
3\left(\frac{\dot{a}}{a}\right)^2  &=
        - \frac12\frac{\Upsilon}{a^6} + \frac{3}{2} \frac{\Xi}{\beta^4a^2} + \Lambda + \varepsilon\\
\label{eq:Friedman2}
2\frac{\ddot{a}}{a} + \left(\frac{\dot{a}}{a}\right)^2 &=
        + \frac12\frac{\Upsilon}{a^6} +  \frac12 \frac{\Xi}{\beta^4a^2} + \Lambda - k \varepsilon
\end{align}
\end{subequations}
This leads to interesting consequences, like a big bounce instead of a big bang for any value $\Upsilon>0$, as considered in \cite{Schmelzer}, which are outside the domain of interest here. But, whatever the values we have to use for $\Xi$, $\Upsilon$ from cosmology, it seems clear that they become completely irrelevant for Solar system observations, for the same reason that the value of the cosmological constant $\Lambda$ is irrelevant here: The dominant terms for Solar system size effects will be those depending on derivatives of the metric. 

For the parameter $\Upsilon$, it is not only the consideration of cosmological vs. local effects which suggests that it is too small to have any physical consequences. It should be small on cosmological scale today too, because its influence on cosmology decreases with $a(\tau)$ as $\Upsilon a^{-6}$. But $a(\tau)$ has increased a lot during the expansion of the universe. So, while we clearly have no lower bound, we can be sure that $\Upsilon$ has to be extremely small, far too small to have any influence on stellar dynamics today. 

So what is wrong with this qualitative argument? The point is that, different from Einstein's cosmological term, the additional terms of GLET also depend on the preferred coordinates, moreover, on derivatives of them. So, the argument works only if it is unproblematic to use local harmonic coordinates. This is correct in usual stellar systems. But it becomes very problematic near the horizon of a black hole. Near the horizon, the harmonic coordinates become singular. And, because of this singularity, even if multiplied with an extremely small value of $\Upsilon$, the additional term becomes important. 

And, indeed, for $\Upsilon>0$, however small, the gravitational collapse, which in GR leads to a black hole, looks completely different. As already mentioned, we can reuse here the results of RTG. So we can refer the reader to sec. 11 of \cite{Logunov1} (see also \cite{Logunov2}, \cite{GershteinBH}, \cite{GershteinBHBB}, \cite{GershteinBHBB2}, \cite{VlasovBH}). The gravitational collapse stops shortly before horizon formation, and, as the result of the collapse, a ``frozen star'' with size slightly greater than its Schwarzschild horizon remains. The difference exists only in a very small environment of the Schwarzschild radius. Outside this environment, the GR solution is a good approximation. Therefore in the GR limit $\Upsilon\to +0$ recovers only the outer part of the Schwarzschild solution.

This effect appears also in other theories of gravity with non-zero graviton mass \cite{Grishchuk02}. Their common property is that the mass term depends on the flat background and therefore is able to feel locally that time dilation relative to the background becomes extremal. 

Here, \cite{NieuwenhuizenBH} gives (for RTG) a lower bound $z \gtrsim 10^{23} M_\odot/M$.

\subsection{The observational argument of Broderick and Narayan}

In \cite{BN1,BN2,BN3}, Broderick and Narayan claim that observations of black hole candidates, in particular ``of Sagittarius $A^*$ (Sgr $A^*$), the supermassive black hole at the center of the Milky Way, all but require the existence of a horizon'' \cite{BN3}. In fact, if there is a surface, in principle it should be possible to observe surface radiation. In \cite{BN3} Broderick et al claim to ``show that for such surface emission to remain undetected would
require unphysically large radiative efficiencies; greater than 99.6\%, as compared to the typical efficiencies in AGN of 10\% \ldots Most importantly, we do this using only observable
quantities, ensuring that these limits are largely independent of the gravitational theory employed. \ldots Our results hold for any geometric gravitational theory admitting stationary solutions.''

Now, GLET is a theory of this type. If these claims would be correct, GLET would be falsified by observation. So, there is a reason to take a close look at these arguments. Of course, they also have to rely on some physical assumptions: Their ``\ldots argument depends critically upon three underlying, physically well motivated assumptions: (1) Sgr $A^*$ is accretion powered, (2) has reached steady state and (3) compact surfaces are approximate blackbodies.'' They ``wish to emphasize that relaxing any of these would require fundamental alterations to presently well understood physics (such as microscopic physics at pedestrian, and laboratory accessible, densities, temperatures and magnetic field strengths).'' Nevertheless, they discuss each in detail, describing how these are justified in the context of Sgr $A^*$.

It appears that we do not have to question assumptions (1) and (2), because the assumption which is the critical one is the steady state assumption. 

Now, the steady state assumption is well justified for usual compact objects, in particular also such objects like neutron stars. Infalling matter usually has some mass as well as kinetic energy. If we ignore possible transformations of mass into energy, the infalling mass increases the mass of the star, and the kinetic energy the temperature of the star. On the other hand, temperature decreases because of thermal radiation. The steady state assumption is, roughly speaking, that these effects are in equilibrium, thus, the average kinetic energy of infalling matter is radiated away as thermal radiation. 

In this case, it would be, indeed, quite clear that one could distinguish a frozen star, with a surface in steady state, from a black hole.  All one has to do is to estimate the amount of infalling matter. For the infalling matter, one can apply reasonable models based on known and well-understood physics. For the type of outcoming radiation, there is also a quite reasonable model -- it has to be a blackbody radiation. If the infalling matter hits the surface, it is also quite clear that most of its energy is gravitational binding energy transformed into kinetic energy of the infalling body, thus, has to be radiated away. So one can compute the amount of blackbody radiation which would have to come out, if the black hole candidate is in a steady state. Once we do not see this -- and, following \cite{BN3}, we don't -- the case seems closed. 

But it is not. What can be, and has to be, questioned is the steady state assumption.

To see why it may be reasonable to expect that the steady state has not yet been reached, it seems useful to have a look at the surface temperature, as measured by an observer on the surface, which would be necessary to create the steady state outgoing radiation.  

Here we have to take into account that the outgoing radiation is highly redshifted, by a redshift factor $z$ defined (in GR approximation) by
\begin{equation}
(1+z) = \left(1-\frac{R_S}{R}\right)^{-\frac12},
\end{equation}

Second, only a small part of the surface radiation reaches infinity. Most of it simply falls back on the surface because of gravitational lensing. The part which is able to reach infinity is (in GR approximation) defined by
\begin{equation}\label{radiation}
 \frac{d\Omega}{2\pi} = 1-\left[1-\frac{27}{4}\frac{1-R_S/R}{(R/R_S)^2}\right]^\frac12.
\end{equation}
which gives $z^{-2}$ for large $z$ \cite{BHradiation}. The observable flux at distance $D$ of a black hole with surface temperature $T$ for frequency $\nu'$ in the surface frame will be
\begin{equation}
 F_{\nu'} =\frac{2h\nu'^3}{c^2}\frac{e^{-h\nu'/kT}}{1-e^{-h\nu'/kT}}\frac{\pi R^2}{D^2} \frac{d\Omega}{2\pi}.
\end{equation} 
The flux visible in infinity decreases by a factor $(1+z)^{-1}$ because of time dilation. The frequency in the surface frame is $\nu'=(1+z)\nu$. This gives for large $z$
\begin{equation}
 F_\nu = \left(\frac{kT}{1+z}\right)^3 F(\frac{h(1+z)\nu}{kT})\frac{\pi R_S^2}{D^2}\frac{27}{8}^{-3}
\end{equation} 
with
\begin{equation}
 F(\lambda) = \frac{2\lambda^3}{h^2c^2} \frac{e^{-\lambda}}{1-e^{-\lambda}} \le F_{max}
\end{equation}
being a bounded function. Thus, for sufficiently small $\frac{kT}{1+z}$ surface radiation will become invisible. Thus, we would need extremely large surface temperatures $T_{surf}\sim z$ to be able to see the surface radiation at large distances. 

In principle the heating by infalling matter could provide sufficient energy to reach such temperatures. To decide if this happens for a given $z$, we would need a hypothesis about the thermal capacity of the material of the frozen star at extremely large temperatures $T\sim z$. To exclude a surface completely, we would need such a hypothesis even for $T\to\infty$.\footnote{Moreover, we would also have to assume the validity of the EEP even for such extremal redshifts.} We obviously cannot have reliable data in this limit.

To justify the steady state assumption, in \cite{BN3} two characteristic time scales are compared: ``The dynamical timescale of Sgr $A^*$ depends upon the nature of the object. Nevertheless, we may expect that it is comparable to the dynamical timescale of the corresponding black hole, $GM/c^3\simeq 20$ s. This is supported by the $\sim 10$ min variability observed in the NIR and X-rays, presumed to be associated with material orbiting nearby. Both of these are much, much shorter than the estimated age of Sgr $A^*$, $10$ Gyr, making it natural to assume that it has sufficient time to reach steady state.'' So they conclude that ``even surfaces with extraordinary redshifts ($z \lesssim 10^{15}$) will have reached steady state''.

But even for Sgr $A^*$ with mass $M = 4.5 \pm  0.4 \cdot 10^6 M_\odot$ this argument does not give a steady state even for the minimal surface redshift $z \gtrsim 10^{23} M_\odot/M$ given in \cite{NieuwenhuizenBH}. Thus, the minimal surface redshift which is, because of cosmological upper limits for $\Upsilon$,  required, is already large enough to make it impossible to apply the steady state assumption. 

In a similar way it seems extremely unlikely that the impact of infalling matter leads to observable effects. In principle, an ideal point-like elastic infalling body would bounce back from the surface of a frozen star \cite{Logunov1}, \cite{VlasovBH}. But this idealization fails already for comets falling on the Earth. For the frozen stars, it follows from \eqref{radiation} that even a minor scattering away from the vertical direction is sufficient to prevent even an ideal elastic body to reach infinity. Thus, even if, because of energy conservation, an infalling ideally elastic body has enough energy to go back to infinity, it seems extremely unlikely that this leads to observable effects different from those predicted by GR.

Thus, the existence of a surface for black hole candidates does not endanger the viability of our theory even if the surface remains invisible. Instead, the central black hole singularity, which is unavoidable in GR, provides a strong conceptual argument against GR.

\section{General character of the argument}

What can be seen here, with the example of GLET, is not only a counterexample of a viable theory with stable stars instead of black holes. It seems worth to be noted that similar effects appear also in other theories of gravity. In particular, it seems to be a common property of theories of massive gravity.

For such a theory to follow the same scheme, one does not need that much. To see this, we only have to consider the substantial points necessary to obtain the effect. All one needs is some small non-covariant term in the equations which prefers a certain (for simplicity flat) background.  Then, all we need is that the preferred coordinates of this background become invalid near the black hole horizon, similar to the harmonic time $\t$ in GLET, which becomes infinite at the horizon. Then, however small the background-dependent term, it may become infinite where the background coordinate becomes infinite. So one can assume that this leads to a regularization, so that the preferred coordinate does not become infinite. So, if the background coordinates are sufficiently close to harmonic coordinates, which seems not unreasonable, so that they degenerate similarly near horizon formation, one can expect that in this theory no horizon appears. But even if there remains a horizon, the situation has to change in an essential way, to prevent the infinity of the preferred coordinates. 

Then, the smaller the additional, non-covariant term, the larger the domain of applicability of the GR approximation. One can expect that, multiplying the additional term with a small enough number, we can obtain a theory where everything remains like in GR up to an arbitrary large surface time dilation, and only then it starts to differ from GR -- but then it will differ essentially.  

One should not think that it is the horizon itself which causes the problem. The problem is caused by the singularity of the preferred coordinates of the background.  we can expect that close to the horizon formation something chances in a very essential way, probably in such a radical way that no horizon appears.

\subsection{The metaphysics of general relativity}

In this context it seem important to consider the problems of the metaphysics of the spacetime interpretation of general relativity.  

The first objection against the very notion of GR metaphysics is the claim that there is no such animal. All GR cares about are observables. But really?  There is agreement about the fact that ``proper time'' defines the results of clock measurements. Or, in neutral words, proper time is simply clock time. But is clock time everything?  

What about the flow of time we experience? As the result of relativistic time dilation, one thing is clear: We cannot measure it with clocks. I have a clock, you have a clock, at our first meeting, we start them showing the same time. At our second meeting, they show different time. The difference is minimal, but conceptually this does not matter -- the result is the same: The clocks do not measure the flow of time.  

But, once there is nothing in GR which measures the flow of time, maybe we should completely forget about this notion? Then there simply is no such thing as a flow of time. Unfortunately, this is not a very reasonable choice. First of all, because we simply experience the flow of time. Everything changes, that's fact. Everybody observes it. And, even more, there is a connection between the flow of time and clock time, a qualitative one as well as a quantitative: The qualitative one is that time flows in the direction of increasing clock time. And the quantitative one is that clock time works nicely as an approximate description of the flow of time. 

So, a theory which bans the flow of time from physics is simply not complete. I'm not a worldline in some four-dimensional thing named spacetime. There is more, point. At the current moment, I may be located at some point on this worldline. Without this location on my worldline, something important is missed. 

What can one do to meet, on the one hand, the positivistic ideal of a physical theory which describes only objective, exactly measurable things, but, on the other hand, the ideal of a complete theory, which does not exclude completely such an obviously experienced things as the flow of time?  One solution is to fall back to solipsism. Its only me who experiences the flow of time, other people can tell me something about their experiences, but they may be simply robots with no real experiences. So, all one needs is to define the flow of time on a single worldline, the own worldline. And, for this purpose, clock time works nicely. 

One may hide the solipsistic character of this interpretation. Other people also experience a flow of time, no problem. Their flow of time is described by the flow of proper time along their worldline. Let's name this the multisolipsistic interpretation. 

This is the point where the simple twin experiment becomes a twin paradox. There is, last but not least, nothing strange in the fact that clocks are not as accurate as we would prefer them. There is always some influence of the environment. So if I leave one clock at home, and take one with me, there is nothing paradoxical if they show, later, different time -- they have been in different environments. So if I meet, in the twin experiment, my twin later, after his journey, there is nothing problematic with the fact that our clocks show different numbers. 

But, according to the multisolipsistic interpretation, I do not really meet the same twin, but something different -- a sort of empty place on the worldline of my twin, while my twin is really on a different place on his worldline. It may be, possibly, occupied by another clon of my twin, who follows the same worldline. So, or we have to multiply clones without necessity, (violating Ockham's razor), or we have to accept that we sometimes communicate with empty places. But in the last case, it seems more reasonable to go back to pure solipsism. 

Or we have to accept that the flow of time on my worldline is not accurately described by my own clock time. So, if I meet my twin later, we have experienced the same objective flow of time. But, once our clocks show different clock time, this same flow of time is not described accurately, but only approximately, by clock time.

In the last case, there is also a quite simple object which can be used to describe the flow of time, namely a global time-like coordinate. The condition that it is time-like meets the observation that clock time increases with the flow of time. What is missed is that proper time is, at least locally and for small velocities, an accurate approximation for the flow of time. 

So one can conclude that there is no consistent metaphysics of the flow of time in general relativity. There are four variants of it -- no flow of time at all, solipsistic flow of time, the multi-solipsistic variant of it, and the flow of time as an otherwise unspecified time-like coordinate. All of them are more or less unsatisfatory. 

The positivistic prejudice against metaphysics allows to hide this unsatisfactory situation. It is simply not appropriate for a physicist to discuss such things at all. And, whenever somebody starts to discuss such things, one can fallback to the ``no flow of time'' position.

A satisfactory interpretation of the flow of time in general relativity would be, instead, the following:  There is an objective flow of time, and this flow of time is defined by a global time-like coordinate $\t$. There should be an additional equation, which describes the flow of time relative to clock time, but this additional equation is unknown. Thus, general relativity is simply an incomplete theory.

Given that, from point of view of simplicity, there is a nice candidate for this unknown equation for  $\t$, namely the harmonic condition: In harmonic coordinates, the Einstein equations themself are much simpler than in general coordinates.

But the particular choice of such an equation is, of course, already speculation beyond the scope of general relativity. The point of this section is a different one: Already the consideration of the GR metaphysics themself -- if one seriously considers them, instead of following the positivistic refusal to discuss them at all -- strongly suggests the incompleteness of GR, and the necessity to complete it with an equation for a preferred notion of time. 

All we need for this argument are the following points:
\begin{itemize}
	\item A complete physical theory should describe everything we observe;
	\item We observe a flow of time, which is connected with clock time, and fulfills the following properties:
	\begin{itemize}
		\item Clock time increases with the flow of time;
		\item Clock time locally, and for small velocities, serves as an approximation of the flow of time; 
	\end{itemize}
\end{itemize}
Everything else follows from the evaluation of various proposals to meet these conditions.

\section{Conclusion}

So, the claim made by Broderick at al that observations of black holes all but require horizon formation have to be rejected. 

We have presented here a particular example of a metric theory of gravity, GLET, which is compatible with the observations of black hole candidates considered, but makes the same predictions about them as general relativity. 

We have found that, even in principle, observations of black hole candidates can give only observational upper bounds for one of the parameters of the theory, $\Upsilon>0$. At the current moment, it does not even give stronger upper bounds than those following from cosmological observations. 

The mathematical result itself -- the very existence of a metric theory of gravity with such properties -- has not been the main point of this paper. In fact, given that the equations of GLET are identical with those of RTG, and that it is already known that in RTG the gravitational collapse does not lead to horizon formation, this would have been not that important. 

But one thing is the existence of some theory with equations which have such properties, and another one the existence of a reasonable physical theory. So, it was also necessary to consider the very justification of the theory, as well as known counterarguments. Given that some arguments against RTG seem to be quite serious -- in particular the cosmological observations like accelerated expansion of the universe, but also the questionable nature of the RTG causality condition -- it is important that there is a theory which does not have these problems. The other popular argument against RTG -- the fear of ghost-like instabilities -- has been rejected as unjustified. This has been done in agreement with similar considerations of this problem by other researchers, but from a slightly different point of view, which seems useful to understand why the ``ghost field'' is not as dangerous as one may think looking at the linear approximation. 

So we conclude that there is not only some theory with such properties, but a physically reasonable and viable one, which gives a counterexample.  

Last but not least, let's note that Broderick at al deserve praise for looking at the region near horizon formation: This is the very region where it is likely that alternative theories will make predictions different from GR. Indeed, this is what we have argued about in the last section: The spacetime metaphysics of GR is far away from being consistent, from being able to describe elementary observations about the flow of time in a consistent, complete, and reasonable way. Many problems make it reasonable to guess that GR is only an approximation of  more fundamental theory, and that this more fundamental theory does not have full relativistic symmetry. And, in this case, the horizon formation of black holes seems likely to be a place where non-covariant properties of such a more fundamental theory may become observable.

\begin{appendix}

\section{Derivation of the Lagrangian}\label{derivation}

Given the condensed matter interpretation, it becomes reasonable to ask for an explanation of the relativistic symmetry of the Lagrangian. Surprisingly, there is a nice and simple derivation of the general Lagrangian from simple first principles. In essence, the relativistic symmetry is obtained as a consequence of the ``action equals reaction'' symmetry of the Lagrange formalism.  

We have already used the so-called ``parametrized formalism'' where we have four explicit scalar functions $\x^\a(x)$ for the ``preferred coordinates'' $\x^\a$ in terms of general coordinates $x^\mu$. In particular, the Lagrangian presented in this form is given by \eqref{Scovgamma} and looks covariant if one forgets about the geometric meaning of the $\x^\a(x)$ as preferred coordinates and considers them as usual scalar fields. 

The key of the derivation is the observation that in the parametrized formalism Euler-Lagrange equations for the ``fields'' $\x^\a(x)$, which can be defined as usual by
\begin{equation}\label{EulerLagrangeCoords}
 \frac{\delta S}{\delta \x^\a} = \frac{\partial L}{\partial \x^\a}
- \partial_\mu \frac{\partial L}{\partial \x^\a_{,\mu}}
+ \partial_\mu\partial_\nu \frac{\partial L}{\partial \x^\a_{,\mu\nu}},
-\cdots 
\end{equation}
give valid equations of motion.\footnote{The only difference from the standard proof for fields is that variations $\x^\alpha+\delta \x^\alpha$ may not define a valid system of coordinates. But for small enough $\delta \x^\alpha$ with finite support, which are sufficient for the proof, this problem does not appear.}
Even more, for the translational symmetry $\x^\alpha\to \x^\alpha+c^\alpha$ of the Lagrangian they automatically give Noether conservation laws
\begin{equation}\label{Ncons}
\frac{\delta S}{\delta \x^\alpha} = \partial_\mu {\mathcal{T}^\mu_\alpha}
\end{equation}
with
\begin{equation}\label{EMT}
\mathcal{T}^\mu_\alpha = -\frac{\partial L}{\partial \x^\a_{,\mu}}
+ \partial_\nu \frac{\partial L}{\partial \x^\a_{,\mu\nu}}
-\cdots. 
\end{equation}

Assuming that we have a four-dimensional parametrized field theory with preferred coordinates $\x^\a(x)$, the following postulate is sufficient for the derivation of the general Lagrangian:

\begin{postulate}\label{postulate}
The energy-momentum tensor $\mathcal{T}^\a_\b$ defined by \eqref{EMT} has the following properties:
\begin{enumerate}
\item For some invertible constant matrix $c^{\a\b}$ the tensor $\mathfrak{g}^{\a\b}=c^{\a\gamma}\mathcal{T}_\gamma^\a$ is symmetric.
\item The $\mathfrak{g}^{\a\b}$ are among the independent field variables.
\end{enumerate}
\end{postulate}

Indeed, this postulate gives together with \eqref{Ncons}
\begin{equation}\label{epostulate}
 \frac{\delta S}{\delta \x^\alpha} = c_{\a\b} \partial_\gamma \mathfrak{g}^{\b\gamma}
\end{equation}
In the four-dimensional case, one can identify the variables $\mathfrak{g}^{\a\b}$ with the usual metric variables $g_{\a\b}$ by
\begin{equation}
 \mathfrak{g}^{\a\b} = g^{\a\b} \sqrt{\det(g_{\a\b})}, \qquad
 g_{\a\b} = \mathfrak{g}_{\a\b} \sqrt{\det(\mathfrak{g}^{\a\b})},
\end{equation}
so that \eqref{epostulate} becomes
\begin{equation}
\frac{\delta S}{\delta \x^\alpha} = c_{\a\b} \partial_\gamma (g^{\b\gamma}\sqrt{g}),
\end{equation}
but the variables $\mathfrak{g}^{\a\b}$ may be used as well as the independent variables. We can define now constants $\gamma_{\alpha\beta}$ by
\begin{equation}
c_{\alpha\beta}  = - (8\pi G)^{-1} \gamma_{\alpha\beta}.
\end{equation}
With these $\gamma_{\alpha\beta}$ the action
\begin{equation}\label{L0}
S_0 = - (16\pi G)^{-1} \int \gamma_{\a\b}g^{\mu\nu}\x^\a_{,\mu}\x^\b_{,\nu}\sqrt{-g}d^4x
\end{equation}
defines a particular solution of \eqref{epostulate}. As usual, the general solution $S$ of the inhomogeneous problem \eqref{epostulate} is the sum of this particular solution $S_0$ and the general solution of the corresponding homogeneous problem, which is in our case
\begin{equation}
\frac{\delta (S-S_0)}{\delta \x^\alpha} = 0.
\end{equation}
Thus, $S-S_0$ should not depend on the preferred coordinates $\x^\a$. But this is simply a well-known way to define the most general Lagrangian of GR. So we obtain
\begin{equation}
S - S_0(g_{\a\b},\x^a_{,\mu}) = S_{GR}(g_{\a\b}) + S_{matter}(g_{\a\b}, \varphi_{m})
\end{equation}
where only the term $S_0$ depends on the preferred coordinates $\x^a$, while $S_{GR}$ as well as $S_{matter}$ have to be covariant, and only $S_{matter}$ depends on the matter fields $\varphi_m$. This action may differ from \eqref{Sgamma} because it can contain non-minimal interactions with matter fields in $S_{matter}$ and higher order terms like $R^2$, $R^{\mu\nu}R_{\mu\nu}$ and so on in $S_{GR}$. But our context is that of an effective field theory, so there is no reason to restrict the action to the lowest order terms \eqref{Sgamma}. All higher order terms may be present, they will be simply suppressed in the large distance limit.

Note that this derivation not only derives the gravity part $S_0+S_{GR}$ of the action, but also the main property -- covariance -- of the matter part $S_{matter}$, or, in other words, the Einstein equivalence principle. The main ingredients of this derivation of relativistic symmetry are the action-equals-reaction symmetry of the Lagrange formalism and the choice of the $g_{\a\b}$ (or, equivalently, of the $\mathfrak{g}^{\a\b}$) as independent variables so that the conservation laws do not depend on the matter fields $\varphi_{m}$:
\begin{equation}
\frac{\delta}{\delta \x^\alpha}\frac{\delta }{\delta \varphi_{m}}S = \frac{\delta}{\delta \varphi_{m}}\frac{\delta }{\delta \x^\alpha}S = \frac{\delta}{\delta \varphi_{m}} (c_{\a\mu} \partial_\nu g^{\mu\nu}\sqrt{g}) = 0.
\end{equation}

How restrictive is our postulate \ref{postulate}?  This is an interesting question which we have to leave to future research.

\end{appendix}

\end{document}